# Methods for analyzing surface texture effects of volcanoes with Plinian and subplinian eruptions types: Cases of study Lascar (23° 22' S) and Chaiten (42 ° 50' S), Chile.

**Abstract:** This paper presents a new methodology that provides the analysis of surface texture changes in areas adjacent to the volcano and its impact product of volcanic activity. To do this, algorithms from digital image processing such as the co-occurrence matrix and the wavelet transform are used. These methods are working on images taken by the Landsat satellite platform sensor 5 TM and Landsat 7 ETM + sensor, and implemented with the purpose of evaluating superficial changes that can warn of surface movements of the volcano. The results were evaluated by similarity metrics for grayscale images, and validated in two different scenarios that have the same type of eruption, but differ, essentially, in climate and vegetation. Finally, the proposed algorithm is presented, setting the parameters and constraints for implementation and use.

L. Fernández • G. Álvarez

Universidad de Antofagasta, Ave. Angamos 601, Antofagasta, Chile

e-mail: luis.fernandez@uantof.cl

G. Álvarez

e-mail: gabriel.alvarez@uantof.cl

R. Salinas

Universidad de Santiago de Chile, Ave. Bernardo O'Higgins 3363, Santiago, Chile

e-mail: renato.salinas@usach.cl

# 1 Introduction

In the process of a volcano eruption in general, it is possible to appreciate some spatial deformation of the volcano, also, depending on the type of eruption it is possible to measure the effects of the eruption on the surface (Fernandez L, et al. 2012). Studies have been conducted to evidence of multi deformations pre-eruptive disorders, post-eruptive and assess the impact of the eruption (Ortiz 1996). The methods can be grouped into two: On one side, the direct methods, which are inclinometry, leveling, check, distance using the impedance, etc. On the other side, we have the distance methods, by which measurements can be performed through a remote sensor. In this group, there is a huge range of devices, including laser ranging, photogrammetry, radiometry, GPS, etc.

It has been documented in the literature about surface deformations using satellite images (Whelley et al 2010; Pavez et al., 2006). As for images captured by passive sensors, Landsat have been used primarily for the study of thermal anomalies (Glaze et al 1989a, 1989b; Wooster 2001), morphologic classification (Bailey et al., 2007), evolution of the volcanic dome (Francis and Rothery 1987; Oppenheimer et al., 1993), among others.

Because of the deformation of the surface of the Earth, one can hope detecting these changes with a remote sensor; however, remote sensors depending on your bandwidth and radiometric resolution are not able to detect these changes on the surface or, in other words, the change of the terrain texture (Haralick et al. 1973). Specifically, in the range of visible wavelengths some shape changes do not alter the colour field, in this case the changes go unnoticed, on the other hand, in the range of radar wavelength it is impossible to detect changes in form altering the earth's surface that are parallel to the line of flight of the satellite platform (azimuthal direction).

The texture theme discussed in the literature distinguishes four methods, mainly: statistical, geometric or structural, based on models and based on transforms or signal processing (Tuceryan and Jain 1998; Materka and Strzelecki 1998; Epifanio 2002; Sucar and Gómez 2003; Fernandez A, 2007; Iglesias 2009).

Because of these methodologies, a number of products obtained need to be validated, it is necessary to establish measures that allow a comparison sizing the existence of changes. In the case of grayscale representations, it is common to use similarity metrics (Svedlow et al. 1976), correspondence regions (Marcello et al. 2007) or disparity maps (Colodro et al. 2012).

# 2 Hypothesis

Volcanic eruptions have shaped the surface of the Earth from its early origin. Movement of magma through the volcanic edifice characterizes these processes. This process can occur with deformation of the surrounding surface or it may be overlooked from the geomorphological point of view, however, the impact on the surface of the Earth of volcanic ash, Pyroclastic flows and morphological deformations, depending on the scale, may become evident and measurable by sensors deployed in satellite platforms in almost all ranges of the electromagnetic spectrum. Algorithms such as co-occurrence matrix, wavelet transform, Sobel operator and growth regions have been successful to classify satellite images for their texture. Therefore, if there are changes in the texture of the surface of the Earth in a volcanic eruption process, they should be classifiable and measurable to a scale consistent with the resolution of the sensor. Using Landsat satellite images with algorithms validated in the literature for seeking changes in the texture of the surface of the Earth can prove useful for solving this problem.

## 3 Study Area

This study builds on two very different scenarios from the geographical point of view, but they have in common the type of eruption. The first scenario, identified as *test sector*, is the Lascar volcano located in northern Chile about 70 km SE of San Pedro de Atacama, in the second region of Antofagasta (Fernandez L., et al. 2012). The second scenario, identified as *validation sector*, corresponds to Chaiten volcano in southern Chile located about 10 km to the east of the Gulf Corcovado (Pierson et al. 2013) in the region of Los Lagos.

According to its shape, the first corresponds to a volcano of mixed type composed of two irregular cones and truncated (Fernandez L, et al. 2012), while the second is a caldera type volcano composed of a complex of domes of rhyolite (Pierson et al. 2013).

Regarding the eruptive activity of these volcanoes, Lascar presents a constant activity especially in the period 1989 - 2001, with the eruption of major importance which occurred in April 1993, corresponding to an eruption of subplinian type generating an eruptive column up 25 km high (Fernández L, et al. 2012). Whereas the volcano Chaiten has no constant activity, such that the first historical eruption occurred in May 2008 (Venzke et al. 2009), characterized as subplinian type generating an eruptive column up to 19 km high (Alfano et al. 2012).

The images used for the implementation and development of this work were obtained from the image bank of the National Institute of Special Investigations (INPE), Brazil. From the products available, images captured by the satellite Landsat 5 TM sensor for the test area, and satellite images Landsat 7 ETM + sensor for the validation sector were used, which have a temporary coverage that covers pre period and post events. Checkpoints and radiometric correction were performed according to the methodology proposed by Chander et al. (2009), thereby achieving a common radiometric scale and its conversion to radiance and thus obtaining a common basis for comparison.

## 4 Development

In our research, we plan and develop two scenarios, both in order to test our hypothesis. In particular, we need a strategy to demonstrate a change of background texture to an eruptive event. This process is especially complex when trying to classify a soil obviously created by volcanic activity in its recent geological past with an activity past volcanic immediately as in the case of Lascar (1993), in the Atacama desert of northern Chile. The second scenario is the event of the Chaitén volcano (2008). These scenarios are essentially characterized by their differences in climate and vegetation.

### 4.1 Methodology

The methodology used, contemplates using the Karhunen-Loeve transform to obtain a set of images called principal components to choose the ones that present better contrast. This technique is widely used in the literature (Fernandez L, et al 2012; Gupta et al 2013), to reduce the number of bands having high correlation, such as the TM and ETM + sensor (Chuvieco 2002; Cheng et al. 2006).

The covariance matrix (Gupta et al. 2013) was used to obtain the main components. As for the choice of the component selected, although the greater variability of information occurs in the first three principal components, in our study we used the third component.

### 4.2 Study Rationale

The texture on a satellite image is understood as the spatial distribution of gray tones and their statistics such as uniformity, density, roughness, regularity, linearity, direction, frequency and phase (Passoni 2005), thickness and intensity (Haralick et al. 1973; Gonzalez and Woods 2002). In short, what is sought is to capture these qualities in order to characterize the texture analytically, to mathematically model these spatial interactions (Gomez M, and Salinas 2006). The characterization has been grouped into four methods which are statistical, geometric or structural, based on models and based on transforms or signal processing (Tuceryan and Jain 1998, Materka and Strzelecki 1998; Epifanio 2002; Sucar and Gómez 2003; Fernandez A, 2007; Iglesias 2009).

In our work, with the purpose of demonstrating change in texture, two methodologies are implemented: statistical methods and signal processing. Regarding the first, co-occurrence matrix is used. The second method uses wavelet transform, which has a number of characteristics that lead to its use as are its ability to locate spectral characteristics of a signal in time or space (Materka and Strzselecki 1998), can vary the spatial resolution and representing textures of various natures (Fernandez A, 2007).

The matrix of co-occurrence represents changes in the distribution of intensities or gray levels of an image and whose elements indicate the frequency with which two gray levels, say *i* and *j* are given in the image taking pixels in pairs and separated by a distance *d* according to a given orientation *θ* (Fernandez A, 2007).

$$P_{i,j} = \frac{A_{i,j}}{\sum_{i,j=0}^{n-1} A_{i,j}} \tag{1}$$

Where $P_{i,j}$ is the number of times an event occurs, divided by the total number of possible events, then obtaining a probability matrix; from this it is possible to extract certain statistics that model the texture (Haralick, et al. 1973).

For texture estimation in a satellite image, using these statistics requires the determination of a sliding window in which the matrix of co-occurrence is obtained, therefore the respective descriptor whose value represents the entire window, is assigned to pixel corresponding image textures. In determining the size of the window there are no specific rules, but there exist some considerations:

1. A large window involve considering information about contiguous pixels belonging to another class that interfere in the calculation, while in a small window do not exist sufficient information to show the entire pattern of the texture element (Fernandez A, 2007) .

2. The choice of window size is based on the image resolution, the minimum unit size mapping, properties or characteristics of the classification and the nature of the object of study (Myint 2003).

3. From studies using the co-occurrence matrix, most used windows have been 3x3, 5x5, 7x7, 11x11, 21x21, and even 65x65.

The wavelet is defined as an oscillatory function of small size, which has both spatial and frequency localization simultaneously (Martinez P, 2012).

$$\psi_{s,u}(x) = \frac{1}{\sqrt{s}} \psi\left(\frac{x-u}{s}\right) \tag{2}$$

Where *s* is the scaling factor and *u* is the translation factor (Ruiz et al. 2005).

To apply the Discrete Wavelet Transform (DWT) on images, the most common form uses filter banks, highpass and lowpass (Ruiz et al. 2005). These filters are applied to the rows and then the columns, in each step, a subsampling is performed by a factor of two, (data is halved), then turning the first level of transformation (scaling), breaking the image into a series of sub-images of low frequency and a set of high-frequency detail.

### 4.3 Applications

To obtain preliminary textural products using different methodologies, the following conditions were considered:

1. Co-occurrence matrix: window size of 3x3 pixels to obtain the co-occurrence matrix, with a distance one between pixels. As to the orientation, the orientation of 135 ° is used, being coincident with the orientation in which pyroclastic flows were deposited, derived from the main events of the Lascar volcano, these being northeast and mainly southeast.

2. Discrete Wavelet Transform: in the absence of standardization regarding the use of wavelet families to obtain textures, we choose to implement Daubechies wavelet of order 2.

The products obtained are presented graphically using Matlab colormap.

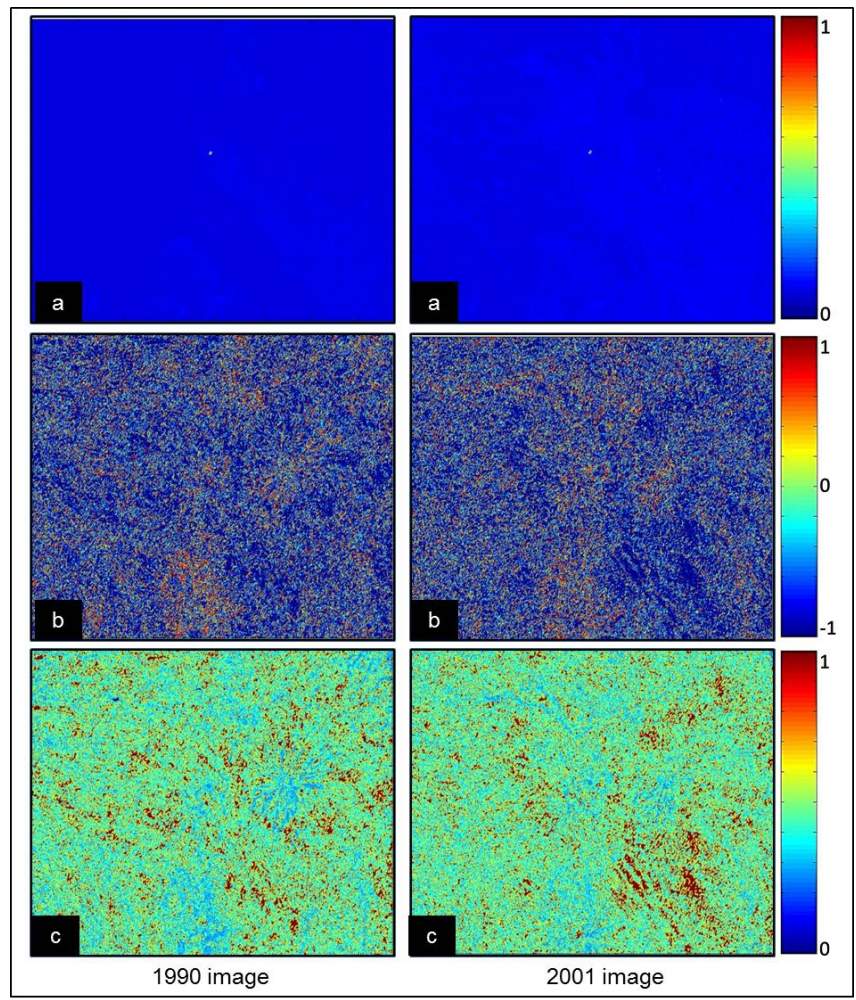

**Fig. 1** Lascar Volcano, co-occurrence matrix.
a) Contrast, b) Correlation, c) Energy.

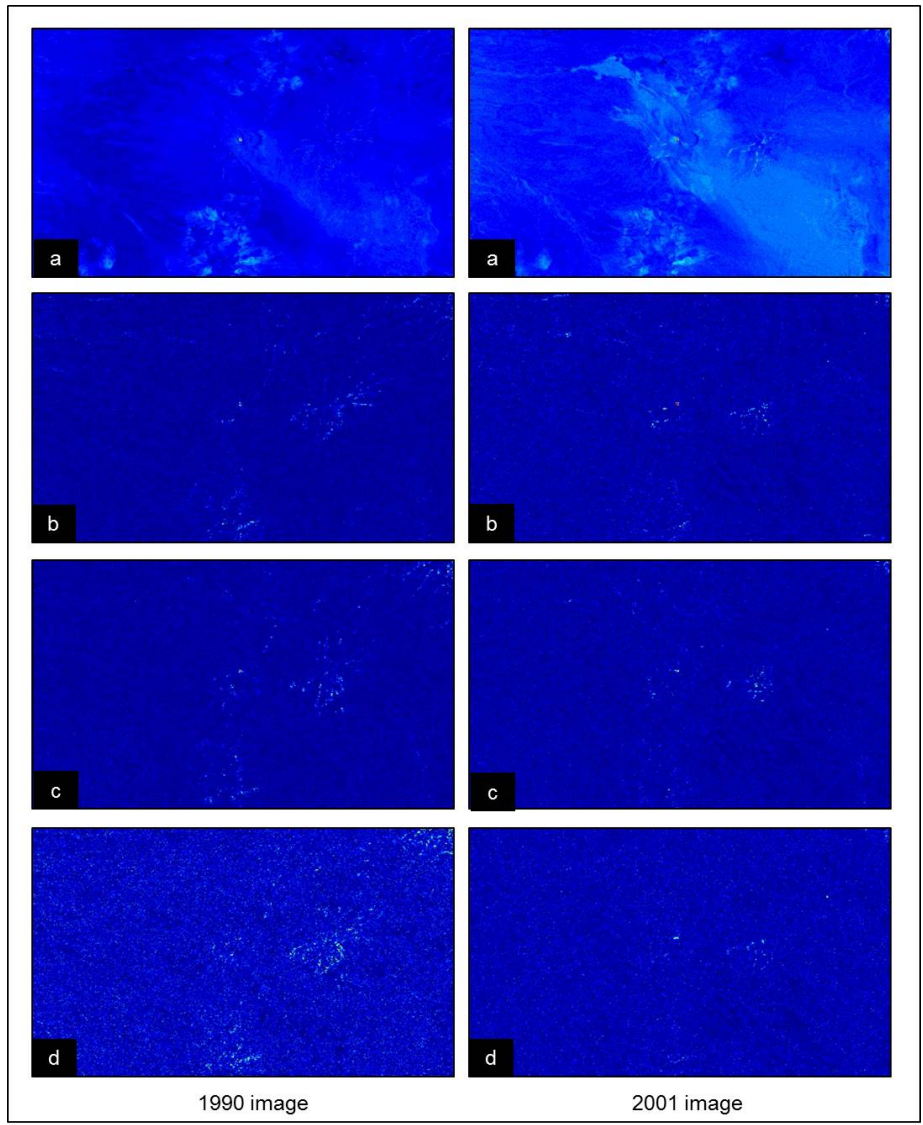

**Fig. 2** Lascar Volcano, Daubechies wavelet coefficients of order 2
a) Approximation, b) Horizontal, c) Vertical, d) Diagonal.

### 4.4 Analysis of the results of texture

**Co-occurrence matrix**: to check the influence that the window size has regarding the results obtained, the sizes of 3x3, 5x5 and 7x7 were analyzed. Fig. 3 shows that increasing the window size, co-occurrence images are obtained with a coarse texture, highlighting in the case of 7x7 the correlation statistics and energy, which define the contour of the reservoir lavas of the northwest slope with respect to the center of the volcano, not so the contrast statistic, which has a fairly regular intensity. The smaller window sizes, 3x3 and 5x5, have more homogeneous textures with little definition of bodies or covers; in particular highlighting the contrast statistic, which highlight the areas defined as change mainly in the 3x3 window.

Focusing the analysis on the statistical test, it is possible to observe from the color bar of Fig. 1, that the concentration value is predominantly at the bottom of it, which shows that the differences between the central pixel and its neighbours in the matrix have no great difference, corresponding to a high number of digital levels of similar value. Statistically, the above is reflected in the co-occurrence matrix where the diagonal closest values will indicate a low-contrast image, while that in a contrasted image there will be a greater number of non-zero elements outside of it.

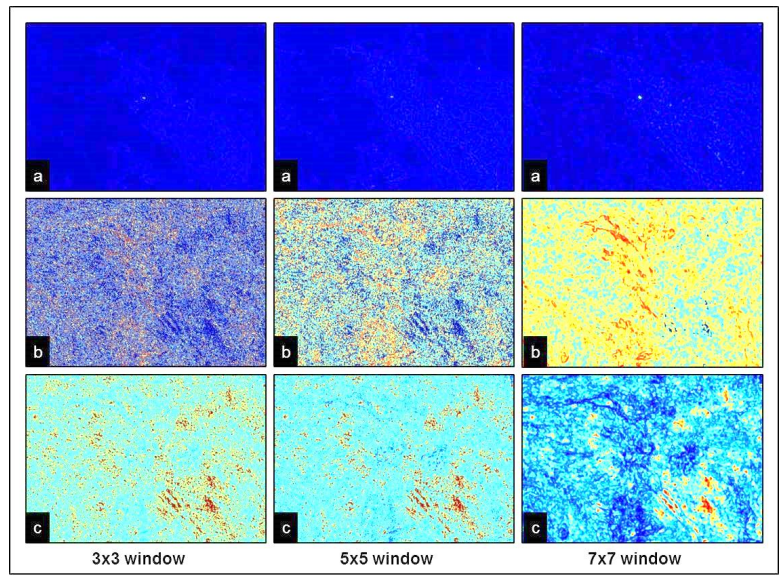

**Fig. 3** Results for different window sizes. 2001 image.

a) Contrast b) Correlation, c) Energy.

**Wavelet Transform**: good results have been obtained using the Coiflet wavelet for the classification of agricultural landscapes by texture (Ruiz et al., 2002); in experimental work using the Vistex data base using the Symlet wavelet (Arivazhagan and Ganesan 2003). In the texture classification of high-resolution images using wavelet Daubechies, Coiflet, Symlet, Battle-Lemarie, Shannon and Meyer (Fernandez a, et al., 2005); using the Haar wavelet for urban classification by texture (Chen et al. 2007) and other using Haar wavelets, Coiflet, Daubachies using the album of Brodatz (Raju et al. 2008), or the topographic analysis strato-volcanos using Meyer wavelet (Gomez C, 2012). In the absence of standardization regarding the use of wavelet families to obtain textures, we chose to implement the wavelet of Daubechies. Firstly, because it has a compact support, i.e. its domain is limited and defined in a closed interval, and there is therefore a number of coefficients and wavelet defined (Misiti et al 1996; Represa 2002; Brown 2011). It is orthogonal, allowing a reconstruction image like the original, in the absence of redundant information (Misiti et al 1996; Pardo 2011). Finally, it does not have an analytic expression; its implementation is performed using digital filters (Misiti et al 1996; Pardo 2011).

For finding the coefficients of Daubechies wavelet of order 2, shown in (Fig. 2), low pass filtering was used to rows and columns. As for coefficients of horizontal, vertical and diagonal detail corresponding to high frequencies, they do not provide nuances of interest when it comes to finding superficial defects.

Change detection aims to detect and account for the transformations occurring in a specific area at a particular time basis for it in determining correspondences or differences. To carry out these metrics, a pattern is set; this pattern is composed of the pixel you want to find its equal and its neighbors in a 3x3-size environment.

**Evaluation:** By implementing similarity metrics with an orientation of zero degrees and zero offset pixels, it is possible to account for the existing surface changes in the period 1990 - 2001. The 1990 image is set as a reference and the 2001 image as the similarity search area.

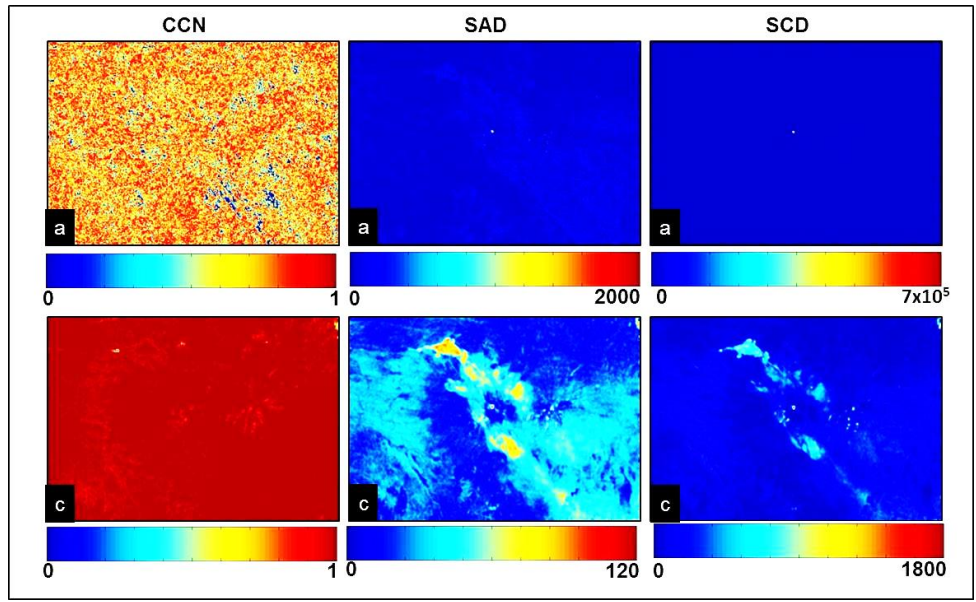

**Fig**. 4 Results for similarity metric.
a) Contrast, b) Approximation coefficients.

Applying the normalized cross-correlation (NCC) to contrast images and approximation images (Fig. 4), we appreciate a marked tendency to the maximum value, according to the color bar, which indicates that there is a great similarity between the standard image and that of search. As for the sum of absolute differences (SAD), it is understood that there is a greater correspondence in those sectors where the sum of the difference is minimal. Fig. 4 shows that for contrast statistics values are obtained close to the minimum value, while for the approximation coefficient, the values are positioned in the middle of the color table, presenting an image with good definition of change areas, therefore superficial defects are easy to observe. The sum of squared differences (SSD) metric means that there is similarity between the reference image and search image when the value obtained of the squared differences is zero. As in the SAD, the products derived from this metric for contrast statistics concentrate on the minimum value. Not so for the approach coefficient presented by the existence of light colors changes. It is understood that the difference between the SAD and the SSD is that the latter gives greater weight to regions with large differences. The results obtained by these metrics depend largely on the window size used to capture differences. A small window (3x3) may not be large enough to capture the variability from one image to another, while a window too large (7x7) is sensitive to the variability of different coverages, giving little definition to the edges.

## 5 Proposed algorithm

The following methodology is proposed, taking into consideration the theoretical aspects discussed in this study and the results obtained, corresponding to a combination of techniques such that, given a set of input images, we detect surface changes.

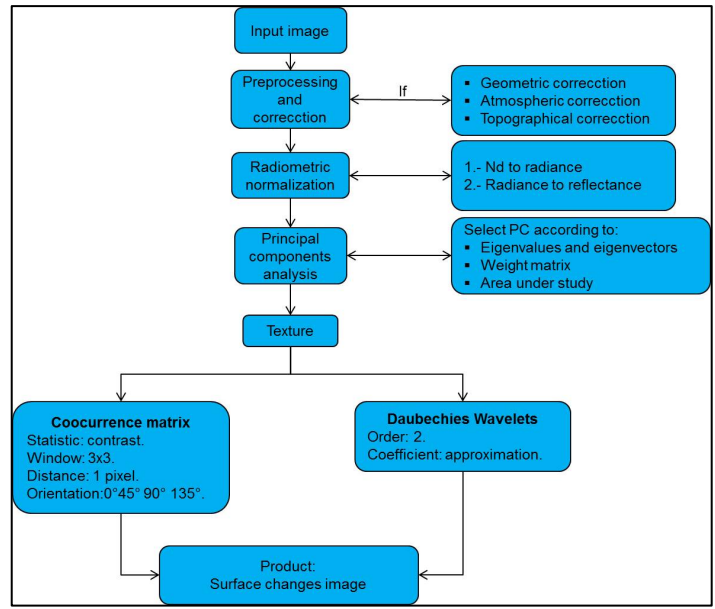

**Fig**. 5 Flowchart for our proposed change detection method

**5.1 Parameterization of the algorithm**

It will be understood as parameterization of the algorithm, the establishment of conditions and restrictions, both the data required as the routines implemented in order to obtain reliable products.

1. Regarding the subject of study. In the particular case of this work, the general objective is to detect surface changes derived from eruptive events in a particular period, therefore, this methodology is recommended for studies of volcanic eruptions of Plinian, Vulcan and / or estrombolian types, whose products are mainly solid or pyroclastic material.

2. With respect to the input images. The input images must have equivalent characteristics in terms of its resolutions. They must be pre-treated to eliminate or reduce any anomalies, as long as these treatments do not cause great alteration of the information.

Being images with different temporality, is recommended a standardized way to achieve a common radiometric scale, for that, the images should be treated by converting digital levels to radiance and then to reflectivity.

The next step is the orthogonal linear transformation, to synthesize information and reduce any possible correlation between bands. This requires getting the principal components. Analysis of these components determine the most appropriate for the study.

3. Regarding the techniques for obtaining texture

For obtaining products that account for surface changes the co-occurrence matrix and discrete wavelet transform are proposed.

Of the statistics that can be obtained from the co-occurrence matrix, contrast is suggested, although others that could also be implemented. This statistic does show the texture of the areas identified as superficial change.

From the results it appears that a window size of 3x3 is ideal to avoid a contamination effect, however, if the images used have a high geometric resolution, it is pertinent an analysis of the effect of any larger window size. Regarding the separation between pixels a distance of one is the recommended value, and in terms of orientation, it is preferable a preliminary analysis of the study area in order to detect the preferred direction in which the various volcanic products are deposited.

Of the existing wavelet families, it is suggested using wavelet Daubechies of order 2, higher orders involve a reduction in the size of the image due to subsampling or decimation process. From the coefficients obtained, the approximation is delivering a unique product for this study.

**5.2 Validation of the methodology**

By way of validating the proposed methodology, results are shown below:

1. The most important eruptive events of the Lascar volcano for the interval 1990-2001, occurred between 1990-1993, 1994-1995 and 2000-2001.

2. Chaiten volcano eruptive event occurred on May 2, 2008.

**First test, Lascar volcano**

Table 1 Lascar Volcano Main Specs

| Location | Latitude: 23°21'55.66"S |
| --- | --- |
|  | Longitude: 67°44'3.93"O |
| Average height | 5592 m.a.s.l. |
| Shape | Volcano composed of mixed type. |
| Type of eruption | Volcanic eruptions, strombolian activity and Plinian eruptions. |
| Activity | Seismic activity, underground noises, ejection of pyroclastic explosions, eruptive column. |

Source: SERNAGEOMIN (Chile).

Table 2 Technical specs of texture. Periods 1990-1993, 1994-1995 and 2000-2001

| Image | Landsat 5 TM |
| --- | --- |
| Principal component | N° 3. |
| Co-ocurrence matrix | Statistic: contrast, Window size: 3x3, Distance: 1 pixel, Orientation: 135° |
| Wavelet | Daubechies Second order, Approximation coefficient. |

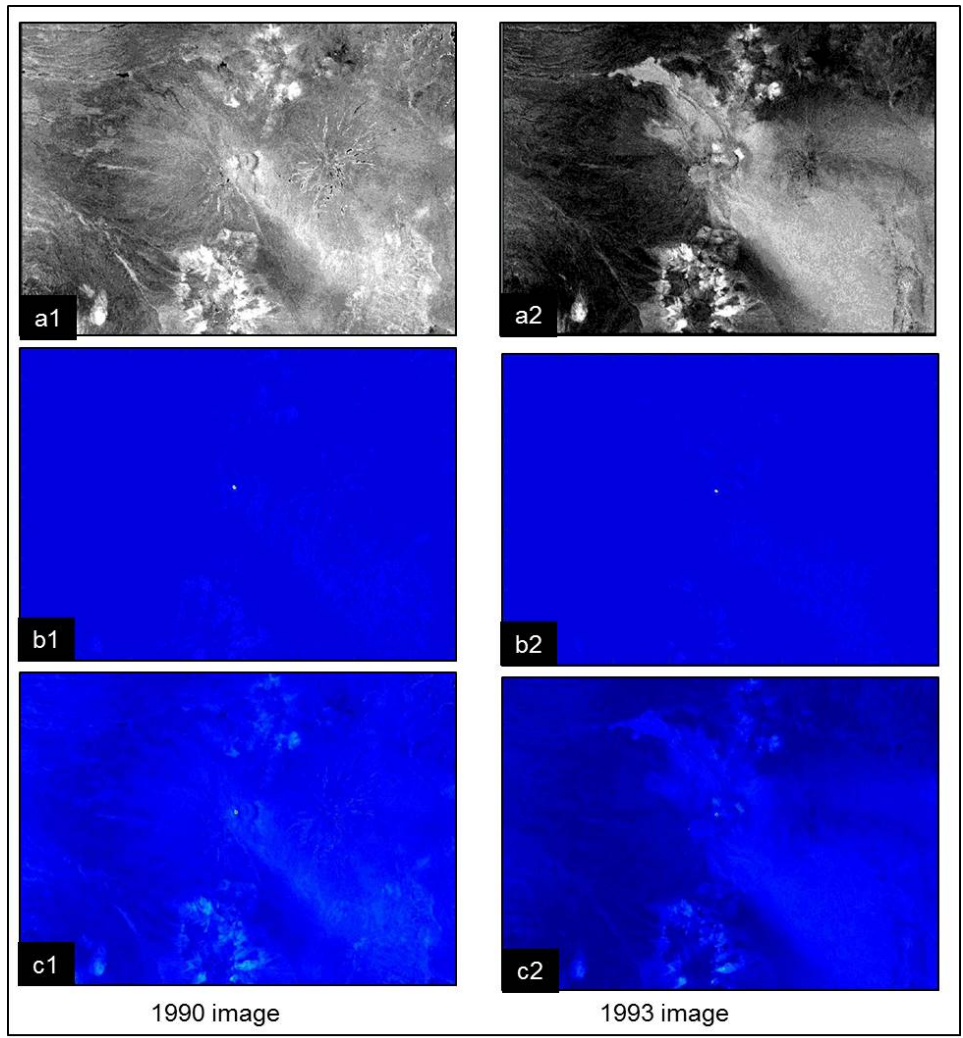

**Fig. 6** Detection of superficial changes, period 1990-1993
a) Principal component, b) Co-occurrence matrix, c) Wavelet.

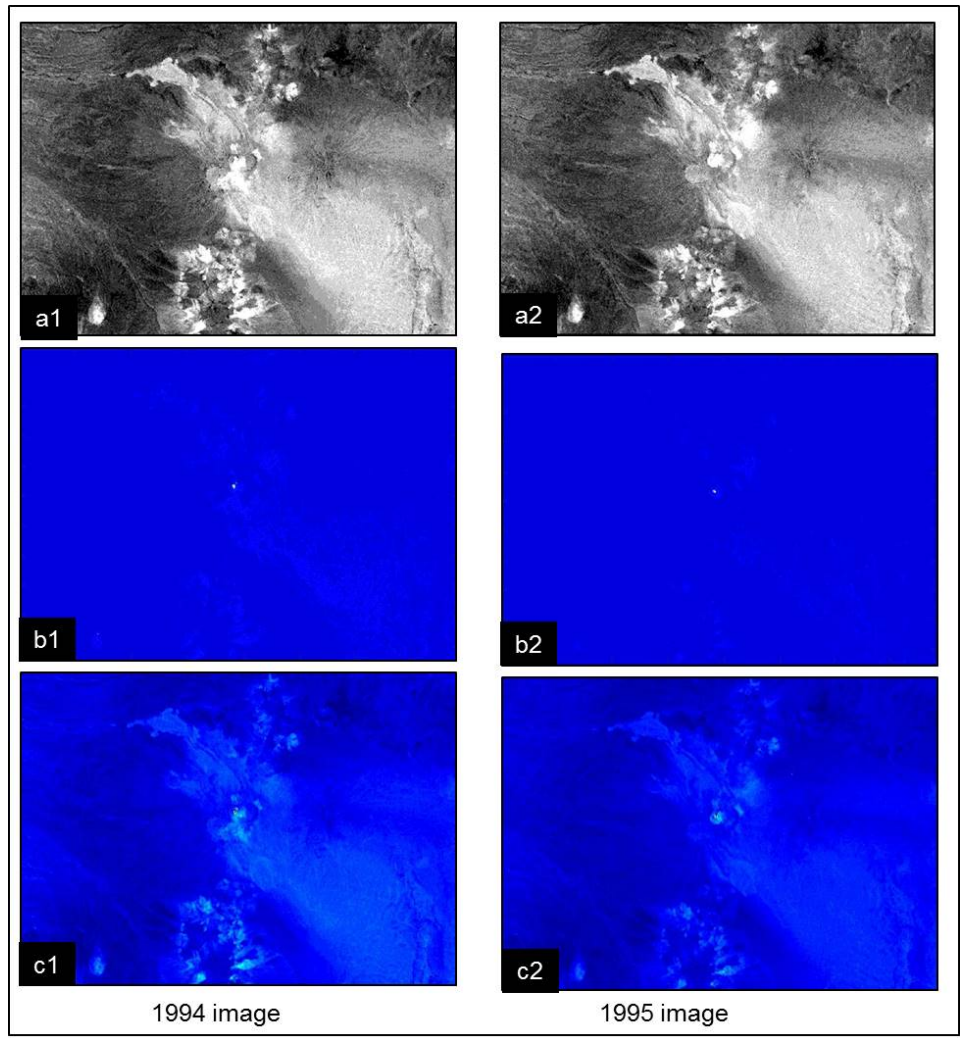

**Fig. 7** Detection of superficial changes, period 1994-1995
a) Principal component, b) Co-occurrence matrix, c) Wavelet.

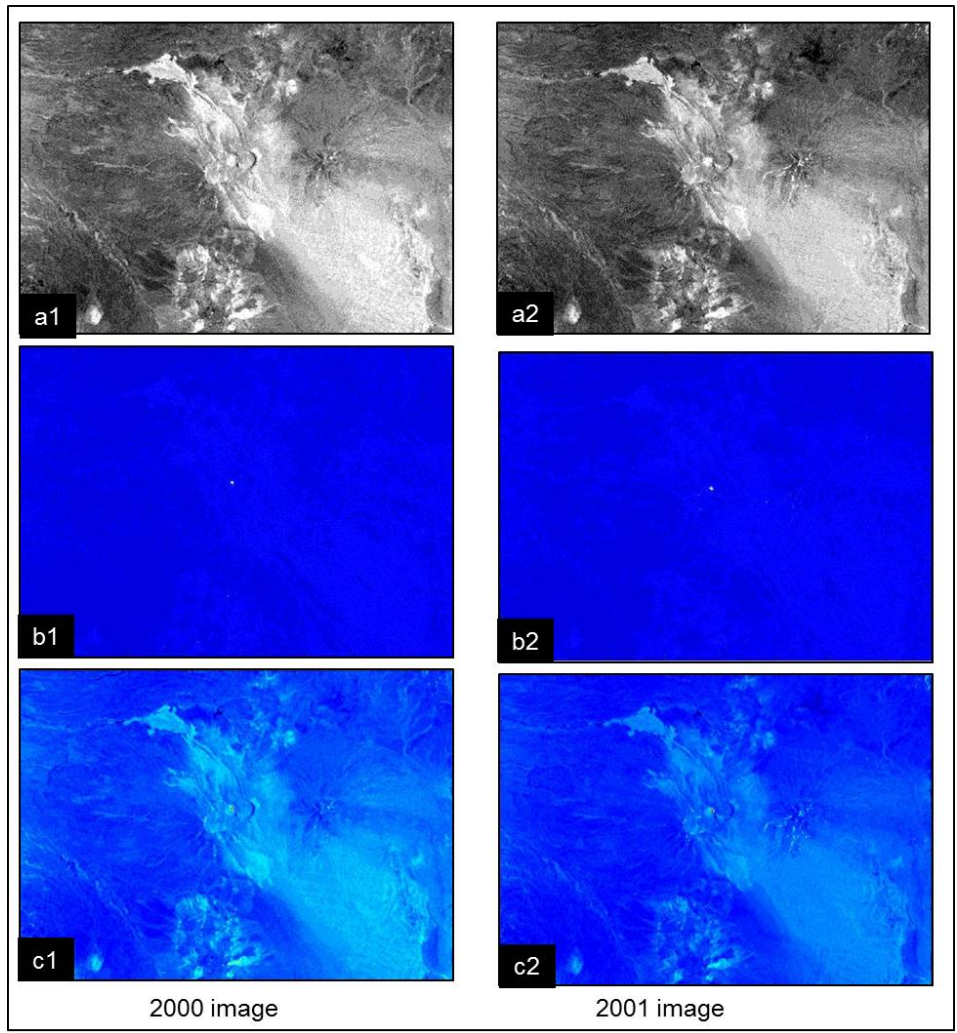

**Fig. 8** Detection of superficial changes, period 2000-2001
a) Principal component, b) Co-occurrence matrix, c) Wavelet.

**Second Test: Chaiten volcano**

**Table 3** Chaiten Volcano Main Specs

| Location | Latitude:42° 50' 8.51" S |
| --- | --- |
|  | Longitude:72° 39' 2.01" O |
| Average height | 950 m.a.s.l |
| Shape | Volcano type caldera |
| Type of eruption | Subplinian. |
| Activity | Seismic activity, pyroclastics ejection, explosion, eruption column, lava dome, lahars. |

**Source:** SERNAGEOMIN (Chile).

Table 2 Technical specs of texture.

| Image | Landsat 7 ETM+ |
|---|---|
| Principal component | N° 3. |
| Co-ocurrence matrix | Statistic: contrast, Window size: 3x3, Distance: 1 pixel. Orientation: 135° |
| Wavelet | Daubechies Second order, Approximation coefficient. |

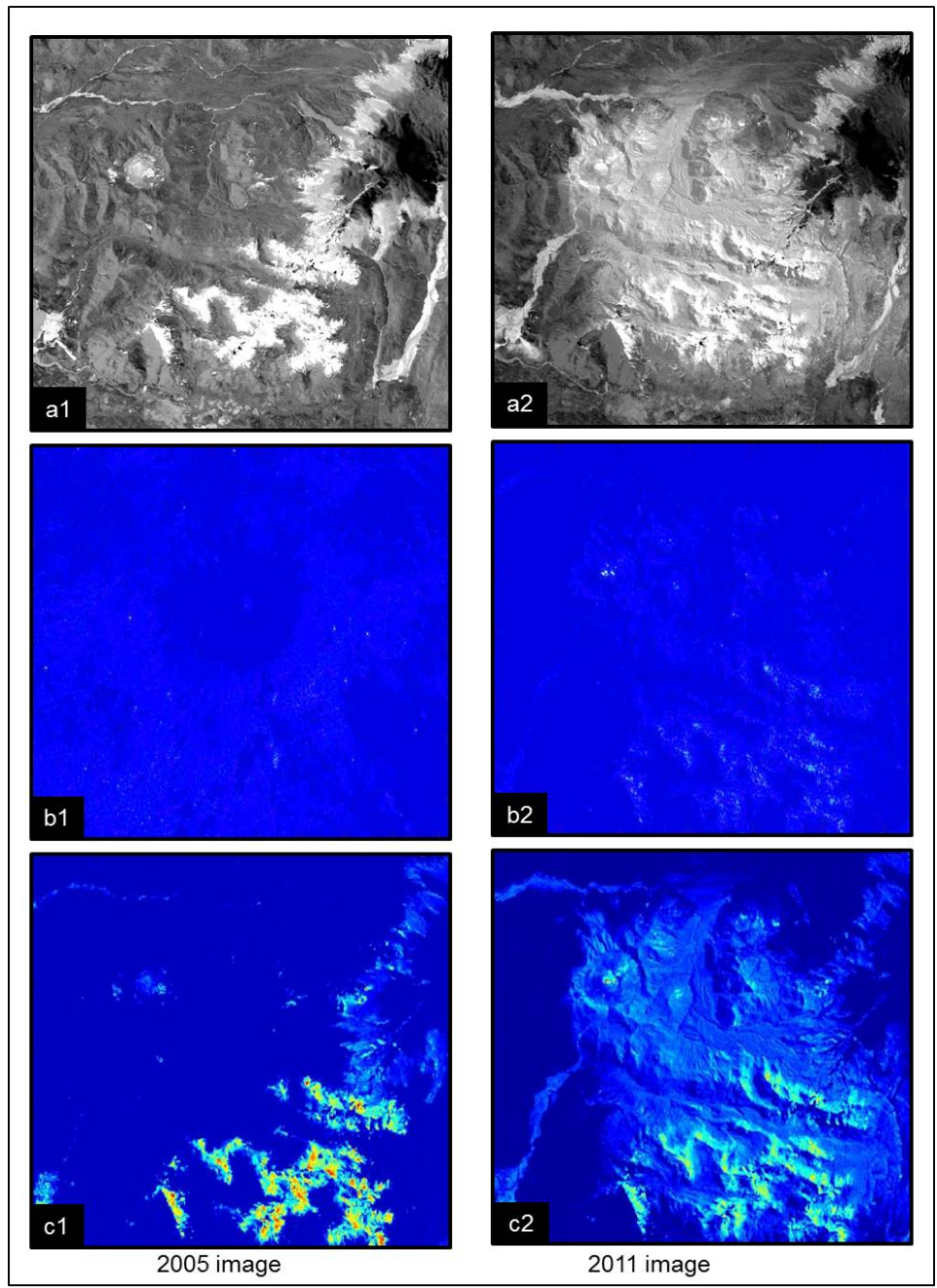

**Fig. 8** Detection of superficial changes, period 2005-2011
a) Principal component, b) Co-occurrence matrix, c) Wavelet.

# 6 Conclusions

As it has been established, change detection aims to detect and account for the changes in a bounded area over a period. Through the study of the theoretical foundations and the analysis of the results obtained by different techniques and methodologies presented, it has been possible to develop an algorithm to detect surface changes resulting from eruptive events.

It has been further demonstrated by different scenarios, that this methodology is valid and therefore can be a contribution to the monitoring and study of possible precursors of an eruptive event if deformation occurs in the volcano during this process.

Finally, through the processes developed and the various tests, the hypothesis established at the beginning of this work should be considered, in view of the results, valid.